# Quantum physics at your fingertips – from paper strips to zippers

### Franziska Greinert*, Malte S. Ubben*

*Technische Universität Braunschweig, Institut für Fachdidaktik der Naturwissenschaften
f.greinert@tu-braunschweig.de

**Abstract**

Quantum physics modeling is technically complex and often non-descriptive. This article presents some approaches how quantum physical ideas can be represented by haptic models. For this purpose, models made from 3D printers, models made from paper strips, and models made from textiles are compared. A novelty is the use of zippers instead of paper strips, which can be easily "cut" and "glued" together. The models have been developed primarily with the aim of conveying and visualizing topological ideas with little basic mathematical knowledge.

1. **Introduction**

In order to meet the growing societal demand for quantum physics education, it is necessary to find suitable representations of quantum physics phenomena that promote understanding of the subject. A major problem is that many ideas, processes, and phenomena in quantum physics cannot be visualized in three-dimensional space for physical reasons. It is therefore a major challenge to provide learners with representations that allow them to understand the complex quantum physics phenomena and that promote as few inappropriate ideas as possible (Ubben, 2020).

One of the biggest problems in learning processes is that representations - such as visualizations - are often perceived by learners as "true to life". In such situations, too much realism is attributed to the representations, preventing learners from abstracting the underlying ideas by holding onto the images (Ubben & Bitzenbauer, 2022; Ubben, 2020).

Because of this problem, one approach is to use abstraction through representations that are well connected to the mathematical descriptions of quantum physics (Schecker et al., 2019). The models presented in this article are based on mathematical topological descriptions proposed by Heusler and Ubben (2019a), among others.

2. **Haptic models**

From a mathematical point of view, quantum states can be represented topologically in Hilbert space. One possible way of visualization is the use of stripes, which can represent objects both in 3D space using Möbius-like bands and in Hilbert space using nodes, without losing much of their mathematical expressiveness (cf. Heusler & Ubben, 2019a). For example, spin states can be represented in this way.

2.1. **Topology and paper strip model**

Let us start with the state $j = 1/2$. This spin state can be represented as a Möbius band of $2\pi$ length with a single rotation in $R^3$. However, considering the corresponding representation in Hilbert space, the band must be extended to a length of $4\pi$. This is done by cutting the Möbius strip lengthwise. The resulting structure is a band with not one, but four turns.

Based on this principle, different topological states in Hilbert space can now be visualized by using the assignment of spin $j = n/2$ to Möbius bands in $R^3$ with $n$ twists. Again, a topological structure in Hilbert space can be revealed by a longitudinal section (see Figure 1). It can be seen that spin states with integer j, i.e. bosonic states, form two knotted copies of themselves, while half-integer j becomes a simple knotted structure.

In practice, however, it is very difficult to reassemble paper strips after cutting them, especially for higher spin numbers such as $j = 5/2$. To solve this problem, zippers can be used.

2.2. **Zippers instead of paper strips**

Zippers make it much easier to "cut" or "rip" the twisted tapes, and it is relatively easy to rejoin the separated tapes.

In addition to the zipper, the simple strip requires some hook-and-loop fasteners and textile adhesive. It is recommended to use two zippers of the same model but different colors, which can be recombined to two two-color zippers. The model identity is important so that the two parts of the different colored zippers fit together and can be used as one two-tone zipper.

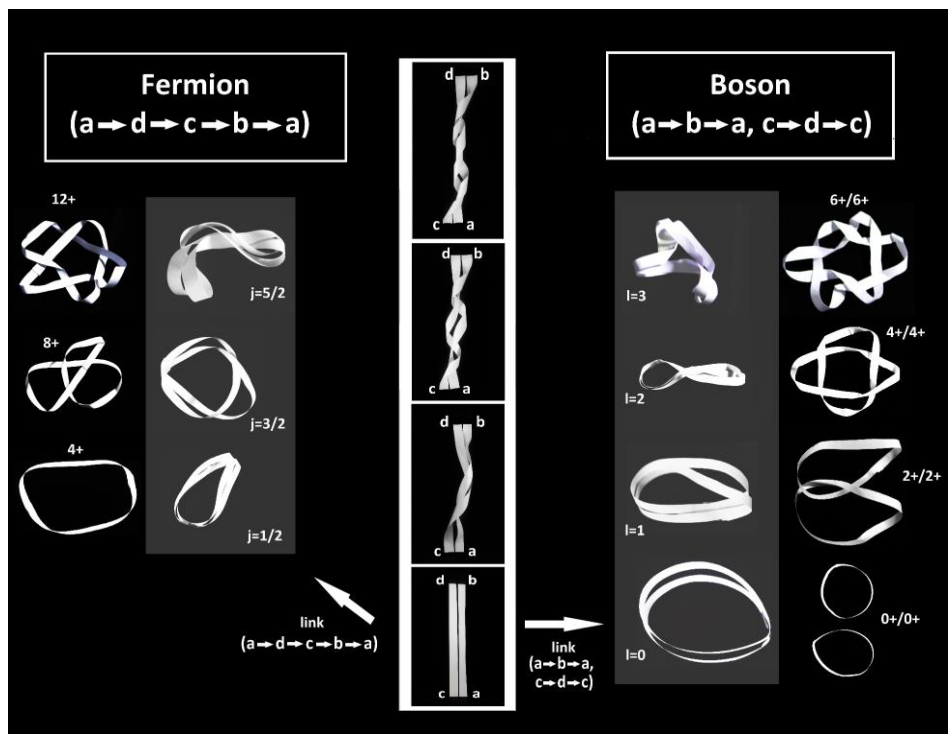

**Fig. 1:** Topological representations of the spin states: Bosonic states duplicate (right) and fermionic states show a singular nodal structure (left) (from Heusler & Ubben, 2019a).

Attach a piece of hook-and-loop fasteners to each end of the zipper on both sides, with the four pieces of hook-and-loop fasteners attached to one end of the zipper and the four pieces of fleece attached to the other end (see Figure 2). Experience has shown that self-adhesive hook-and-loop fasteners does not last long because the adhesive comes off quickly; alternatively, the hook-and-loop pieces can be sewn on.

With such a zipper, spin states can be haptically represented as in Figure 1. As an example, Figure 3 shows the states $j = 1/2$ and $j = 3/2$ (left side, from bottom to top) and $l = 1$ and $l = 3$ (right side, from bottom to top), which are also shown in Figure 1.

### 2.3. Extension: Double strip

The single strip of a zipper can be extended by a second zipper. This double strip then allows the long sides to be connected on both sides, creating a kind of tube or torus without a twist. Again, the use of two different colored zippers of the same model is recommended for better visibility.

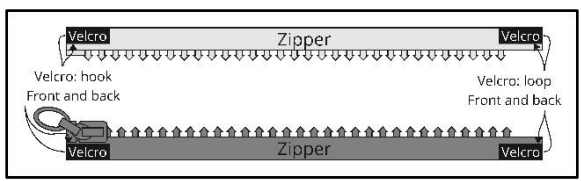

**Fig. 2:** Simple strip: Hook-and-loop fastener pieces are glued to the ends of the two-color zipper.

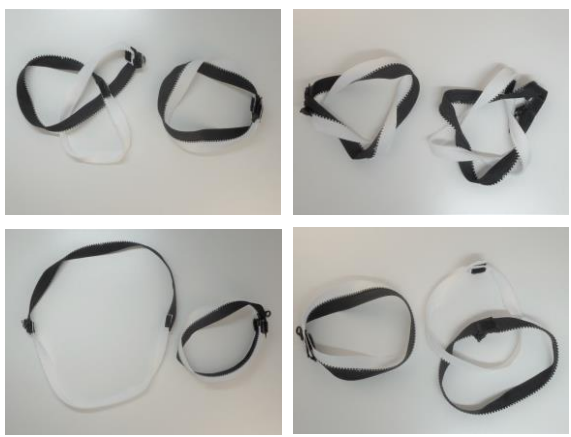

**Fig. 3:** Topological representations of selected spin states with two-color zipper.

A strip of textile of the appropriate color is sewn between each of the two zipper pieces of the same color. At the ends of this strip, hook-and-loop fastener strips are glued as before (see Figure 4). Figure 5 shows the two types, two single strips on the left and the double strip on the right, where two zippers of the same

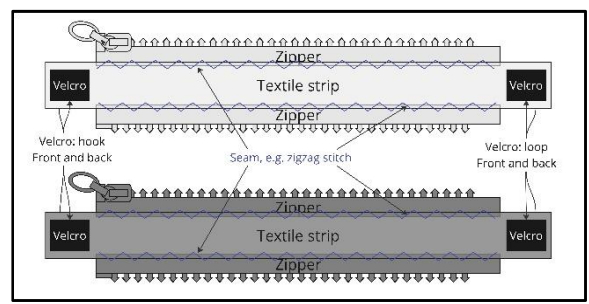

**Fig. 4:** Double strip: strip of textile between the zipper parts and hook-and-loop fasteners at the ends.

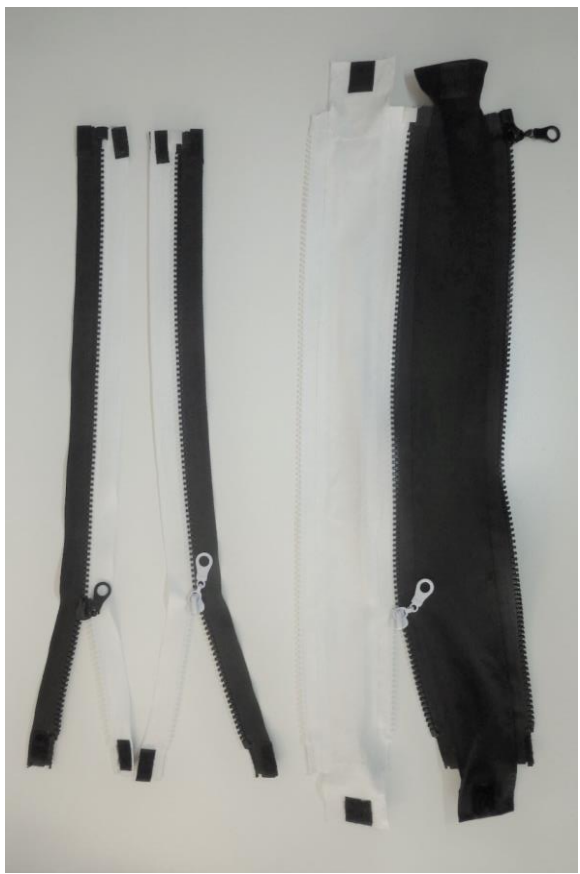

**Fig. 5:** Two single and one double strip, each made of two zippers of the same model in white and black.

model in white and black have been combined to form two-color zippers.

The double strip not only allows you to tear open the zipper and go from $2\pi$ to $4\pi$, but also to further merge the long sides. To illustrate, a Möbius strip can be transformed from a double strip with one zipper closed to a Klein bottle (see Figure 6). In this case, the second zipper cannot be completely closed due to the topological structure of the Klein bottle. A small piece of additional hook-and-loop fastener at the end of this zipper holds it together, and so the typical shape becomes recognizable.

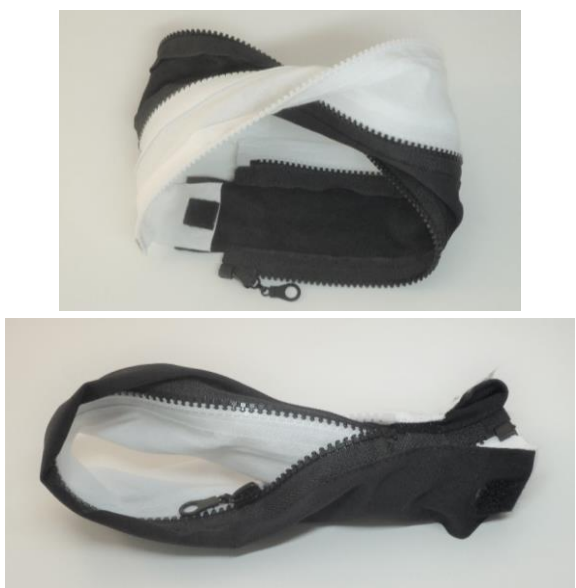

**Fig. 6:** Double strip as a Möbius strip (above), which becomes a Klein's bottle by closing the second zipper.

### 2.4. 3D printing

A visualization of more complex wave functions can be performed using 3D printed spherical surface functions (cf. Ubben & Heusler, 2018, Ubben, 2022). For this purpose, the nodal lines of the spherical surface functions are displayed as red areas within a certain radius (see Figure 7).

It is also possible to visualize spin wave functions in this way. For example, for spin S=1/2, no nodal surface can be seen, only a nodal column. A connection of this representation of a spin with a Möbius strip can be easily made by identifying the nodal column with the twist in the Möbius strip. It is also possible to relate the paper and zipper models to the spherical surface functions.

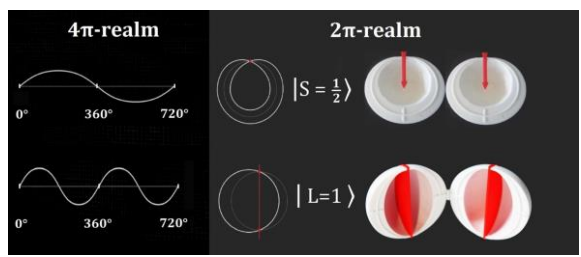

**Fig. 7:** Wave functions as spherical surface functions with knot lines and corresponding 3D printed model (Ubben 2022, p.83).

### 3. Outlook: Further applications

The zipper models presented here could also be used for other topological representations in quantum physics. For example, the paper stripe representation has already been extended to entanglement (Heusler & Ubben, 2019b, 2022, Heusler et al., 2021) and color confinement (Heusler et al., 2021). Again, zippers make the models easier to handle. Especially in color confinement, where three stripes are sometimes used for representation, the use of zippers is appropriate.

Other related mathematical descriptions, such as the generation of a Klein bottle from a Möbius strip, can also be done with the presented zipper models. An implementation of the models in a topologically and/or quantum physics-oriented course may in the future provide more information about the suitability of the haptic representations presented here to support the development of quantum physics knowledge.

## Acknowledgements

We would like to thank Prof. Stefan Heusler for the exciting and constructive discussions and the productive food for thought.